\documentclass[letterpaper,11pt]{article}
\usepackage{amsmath}
\usepackage[dvips]{graphicx}
\begin{document}
\title{Triaxial Analytical Potential-Density Pairs for Galaxies}
\author{Daniel Vogt\thanks{e-mail: dvogt@ime.unicamp.br} and 
Patricio S. Letelier\thanks{e-mail: letelier@ime.unicamp.br} \\
Departamento de Matem\'{a}tica Aplicada-IMECC, Universidade Estadual \\
de Campinas, 13083-970 Campinas, S.\ P., Brazil}
\maketitle
\begin{abstract}
We present two triaxial analytical potential-density pairs that can be viewed as 
generalized versions of the axisymmetric Miyamoto and Nagai and Satoh galactic models. 
These potential-density pairs may be useful models for galaxies with box-shaped bulges.
The resulting mass density distributions are everywhere non-negative and free from 
singularities. 
Also, a few numerically calculated orbits for the Miyamoto and Nagai-like triaxial 
potential are presented.    
\end{abstract}
\section{Introduction}

There are several three-dimensional analytical models in the literature for 
the gravitational field of different types of galaxies and galactic components.  
Jaffe \cite{Jaffe83} and Hernquist \cite{Hern90} discuss 
models for spherical galaxies and bulges. Three-dimensional models for flat galaxies 
were obtained by Miyamoto and Nagai \cite{Miy75} and Satoh \cite{Satoh80}; 
de Zeeuw and Pfenniger \cite{Zeeuw88} presented a sequence of infinite triaxial 
potential-density pairs relevant for galaxies with massive haloes of different shapes. 
Long and Murali \cite{Long92} derived simple potential-density pairs for a prolate and a triaxial bar by
softening a thin needle with a spherical potential and a Miyamoto and Nagai potential, respectively.
See \cite{Binney} for a discussion on other galactic models. There also exist several general relativistic 
models of disks, e.\ g., \cite{Morgan69}--\cite{Vogt2}. A general relativistic version of the Miyamoto and Nagai 
models was studied by Vogt and Letelier \cite{Vogt3}.

Although axial symmetry is a common approximation to many disk galactic models, recent 
statistics of bulges of disk galaxies (S0--Sd) have revealed that nearly half of them are box- and peanut- shaped 
\cite{Luetticke1}. About 4\% of all galaxies with box- and peanut- shaped bulges have so called ``Thick Boxy Bulges'': 
they are box shaped and are large with respect to the diameters of their galaxies \cite{Luetticke2}.
In this work we consider two simple triaxial potential-density pairs whose mass density 
distributions are box-shaped along one axis. The first is constructed by applying on each 
Cartesian coordinate a transformation similar to that used by Miyamoto and Nagai; this is 
done in subsection \ref{sub_sec1}. The second pair is a triaxial generalization of one studied by 
Satoh, and is presented in subsection \ref{sub_sec2}. In section \ref{sec_2} some 
orbits for the Miyamoto and Nagai-like triaxial potential are exhibited. Finally, the results 
are discussed in section \ref{sec_discuss}.   
\section{Triaxial Models for Galaxies} \label{sec_1}

In the following potential-density pairs, the mass-density distribution is obtained directly 
from Poisson equation
\begin{equation} \label{eq_poisson}
\rho= \frac{1}{4\pi G} \left( \Phi_{,xx}+\Phi_{,yy}+\Phi_{,zz} \right) \mbox{,}
\end{equation}
where $\Phi(x,y,z)$ is the gravitational potential. 
\subsection{Triaxial Miyamoto and Nagai-like Model 1} \label{sub_sec1}

We start with the gravitational monopole potential in Cartesian coordinates, 
\begin{equation} \label{eq_phi1}
\Phi=-\frac{Gm}{\sqrt{x^2+y^2+z^2}} \mbox{,}
\end{equation}
and apply the transformations $x\rightarrow a_1+\sqrt{x^2+b_1^2}$, $y\rightarrow a_2+\sqrt{y^2+b_2^2}$
and $z\rightarrow a_3+\sqrt{z^2+b_3^2}$, where the $a_i$, $b_i$ are non-negative constants. Using equation 
(\ref{eq_poisson}), we obtain the following density distribution
\begin{align}
\bar{\rho} &=\frac{1}{4\pi \xi^3\eta^3\chi^3\left[ (\bar{a}_1+\xi)^2+(\bar{a}_2+\eta)^2+(1+\chi)^2\right]^{5/2}}
 \left\{ \bar{b}_1^2\eta^3\chi^3 \left[ (\bar{a}_1+\xi)^2(\bar{a}_1+3\xi) \right. \right. \notag \\
& \left. \left. +\bar{a}_1(\bar{a}_2+\eta)^2+\bar{a}_1(1+\chi)^2 \right] 
+ \bar{b}_2^2\xi^3\chi^3 \left[ (\bar{a}_2+\eta)^2(\bar{a}_2+3\eta)+\bar{a}_2(\bar{a}_1+\xi)^2 \right. \right. \notag \\
& \left. \left. +\bar{a}_2(1+\chi)^2 \right] +\bar{b}_3^2\xi^3\eta^3 \left[ (1+\chi)^2(1+3\chi)+(\bar{a}_1+\xi)^2+
(\bar{a}_2+\eta)^2 \right] \right\} \mbox{,}  \label{eq_rho_my}
\end{align}
where the variables and parameters are rescaled in terms of $a_3$: $\bar{x}=x/a_3$, $\bar{y}=y/a_3$, 
$\bar{z}=z/a_3$, $\bar{a}_i=a_i/a_3$, $\bar{b}_i=b_i/a_3$, $\bar{\rho}=a_3^3\rho/m$, and 
$\xi=\sqrt{\bar{x}^2+\bar{b}_1^2}$, $\eta=\sqrt{\bar{y}^2+\bar{b}_2^2}$ and $\chi=\sqrt{\bar{z}^2+\bar{b}_3^2}$. 

\begin{figure}
\centering
\includegraphics[scale=0.68]{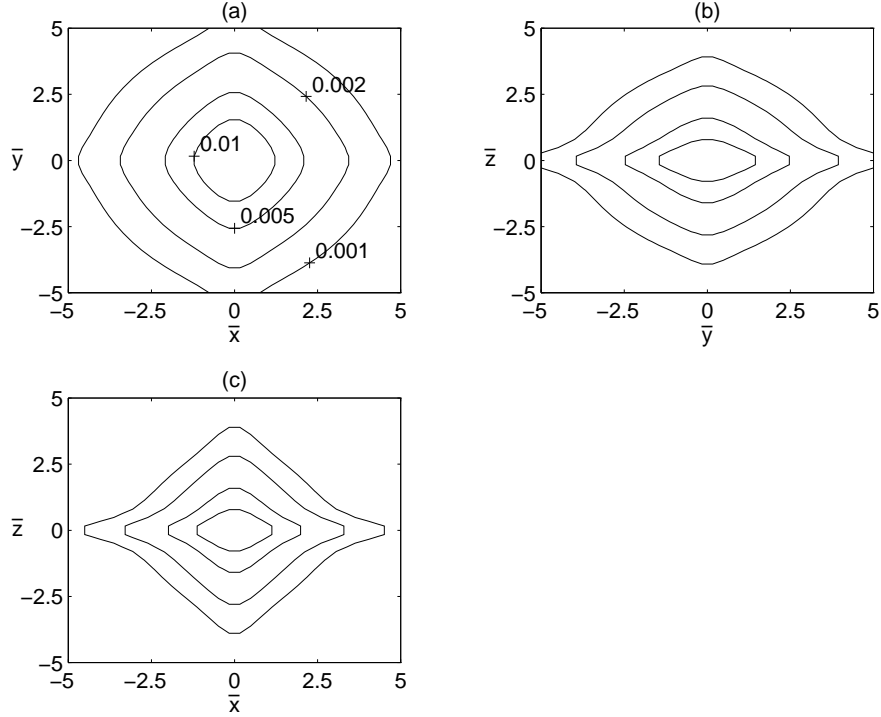}
\caption{Constant density curves of equation (\ref{eq_rho_my}) on the planes (a) $\bar{z}=0$, (b) $\bar{x}=0$ 
and (c) $\bar{y}=0$ with parameters $\bar{a}_1=1.0$, $\bar{b}_1=1.0$, $\bar{a}_2=0.5$, $\bar{b}_2=1.0$ 
and $\bar{b}_3=0.5$. The contour levels in (b) and (c) are the same as shown in (a).} \label{fig1}
\end{figure}
The density distribution equation (\ref{eq_rho_my}) is always non-negative and free from singularities. For 
$\bar{a}_1=\bar{a}_2=0$ the potential-density pair is axisymmetric with respect to the $z$ axis, and in particular if 
$\bar{b}_1=\bar{b}_2=0$ we recover the Miyamoto \& Nagai model 1. Figures \ref{fig1}(a)--(c) show some isodensity 
curves of equation (\ref{eq_rho_my}) on the planes (a) $\bar{z}=0$, (b) $\bar{x}=0$ and (c) $\bar{y}=0$ with 
parameters $\bar{a}_1=1.0$, $\bar{b}_1=1.0$, $\bar{a}_2=0.5$, $\bar{b}_2=1.0$ and $\bar{b}_3=0.5$. From a top  
view, the matter distribution is box-like shaped, whereas from a lateral view matter seems more flattened in a disk-like 
manner. Plotting some isodensity curves with other values of the parameters allows us conclude that larger values of
$\bar{a}_1$ and $\bar{a}_2$ lead to larger deviations from axisymmetry, and the degree of flatness with respect to the 
plane $\bar{z}=0$ depends on $\bar{b}_3$ as in the original  Miyamoto and Nagai models.   
\subsection{A Triaxial Satoh-like Model} \label{sub_sec2}

Satoh \cite{Satoh80} derived a family of three-dimensional axisymmetric mass distributions by flattening the higher 
order Plummer models of order $n$. When $n\rightarrow \infty$ the potential takes a rather simple form 
\begin{equation} \label{eq_phi2}
\Phi=-\frac{Gm}{\left[ x^2+y^2+z^2+a(a+2\sqrt{z^2+b^2}) \right]^{1/2}} \mbox{.}
\end{equation} 
We propose the following triaxial generalization of the above potential:
\begin{multline} \label{eq_phi3}
\Phi=-Gm \left[x^2+a_1(a_1+2\sqrt{x^2+b_1^2})+y^2+a_2(a_2+2\sqrt{y^2+b_2^2})+z^2 \right. \\
\left. +a_3(a_3+2\sqrt{z^2+b_3^2}) \right]^{-1/2} \mbox{.}
\end{multline}
The corresponding mass density distribution follows from equation (\ref{eq_poisson})
\begin{align}
\bar{\rho} &=\frac{1}{4\pi  \xi^3\eta^3\chi^3\left[ (\bar{x}^2+\bar{a}_1(\bar{a}_1+2\xi)+\bar{y}^2+\bar{a}_2(\bar{a}_2+2\eta)+
\bar{z}^2+1+2\chi \right]^{5/2}} \times \notag \\
& \left\{ \bar{a}_1\bar{b}_1^2\eta^3\chi^3\left[ \bar{x}^2+\bar{y}^2+\bar{z}^2+
(\bar{a}_1+2\xi)(\bar{a}_1+3\xi)+\bar{a}_2(\bar{a}_2+2\eta)+1+2\chi \right] \right. \notag \\
& \left. +\bar{a}_2\bar{b}_2^2\xi^3\chi^3\left[ \bar{x}^2+\bar{y}^2+\bar{z}^2+(\bar{a}_2+2\eta)
(\bar{a}_2+3\eta)+\bar{a}_1(\bar{a}_1+2\xi)+1+2\chi \right] \right. \notag \\
& \left. + \bar{b}_3^2\xi^3\eta^3\left[ \bar{x}^2+\bar{y}^2+\bar{z}^2+
(1+2\chi)(1+3\chi)+\bar{a}_1(\bar{a}_1+2\xi)+\bar{a}_2(\bar{a}_2+2\eta) \right] \right\} \mbox{,} \label{eq_rho_sa}
\end{align}
where the variables and parameters were rescaled as in subsection \ref{sub_sec1}.
\begin{figure}
\centering
\includegraphics[scale=0.68]{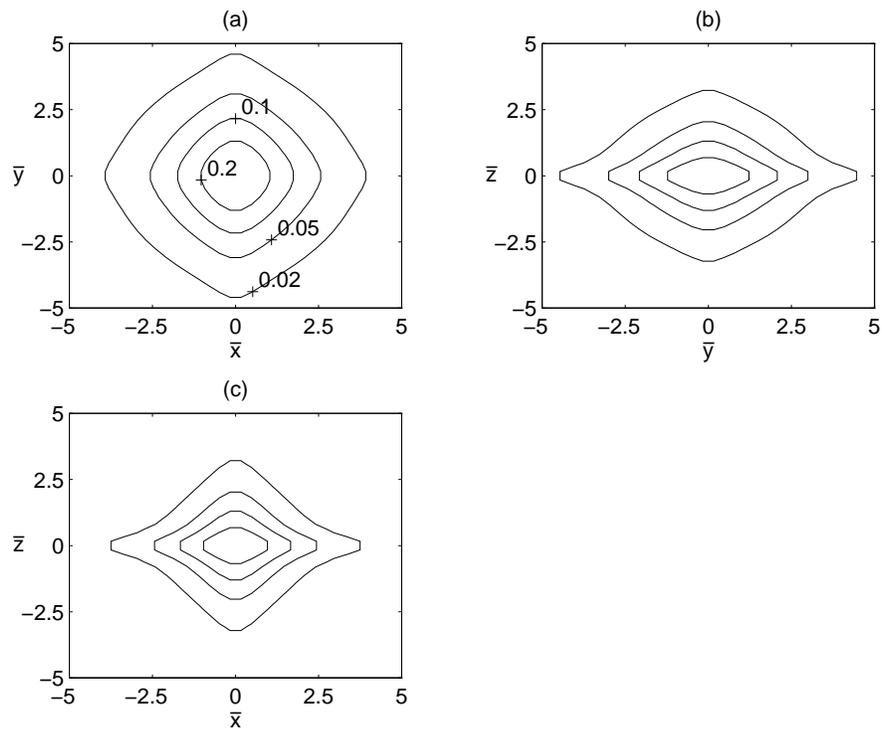}
\caption{Constant density curves of equation (\ref{eq_rho_sa}) on the planes (a) $\bar{z}=0$, (b) $\bar{x}=0$ 
and (c) $\bar{y}=0$ with parameters $\bar{a}_1=1.0$, $\bar{b}_1=1.0$, $\bar{a}_2=0.5$, $\bar{b}_2=1.0$ 
and $\bar{b}_3=0.5$. The contour levels in (b) and (c) are the same as shown in (a).} \label{fig2}
\end{figure} 
The density distribution equation (\ref{eq_rho_sa}) is also non-negative and free from singularities. 
For $\bar{a}_1=\bar{a}_2=0$ we recover the original Satoh model. In figures 
\ref{fig2}(a)--(c) we display some isodensity curves of equation (\ref{eq_rho_sa}) on the planes (a) 
$\bar{z}=0$, (b) $\bar{x}=0$ and (c) $\bar{y}=0$ with 
parameters $\bar{a}_1=1.0$, $\bar{b}_1=1.0$, $\bar{a}_2=0.5$, $\bar{b}_2=1.0$ and $\bar{b}_3=0.5$. 
\section{Orbits in triaxial potential-density pairs} \label{sec_2}

\begin{figure}
\centering
\includegraphics[scale=0.65]{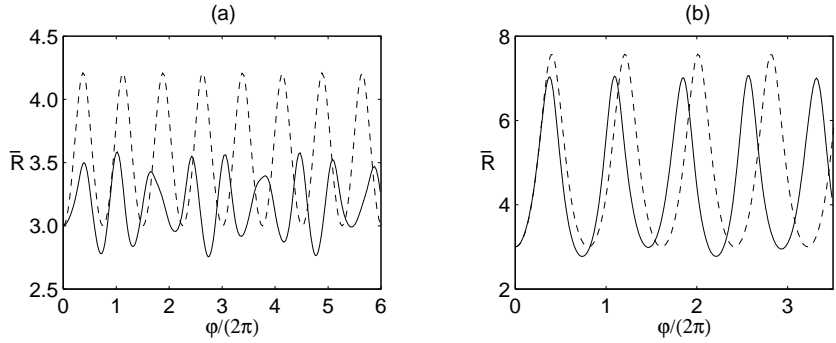}
\caption{Radial coordinate $\bar{R}=R/a_3$ as function of polar coordinate $\varphi$ 
for motion with triaxial potential (solid curves) and with axisymmetric potential (dashed 
curves). Parameters: $\bar{a}_1=1.0$, $\bar{b}_1=1.0$, $\bar{a}_2=0.5$, $\bar{b}_2=1.0$ 
and $\bar{b}_3=0.5$. Initial conditions: $\bar{R}=3.0$, $\dot{\bar{R}}=0$, $\varphi=0$ and 
$\dot{\varphi}$ was calculated for an energy $\bar{E}=-0.15$ in (a) and 
$\bar{E}=-0.10$ in (b).} \label{fig3}
\end{figure}

In this section we briefly discuss some orbits calculated numerically for the triaxial Miyamoto and Nagai-like
potential-density pair presented in subsection \ref{sub_sec1}. We first consider orbits  
on the $z=0$ plane. Since the potential is a function of the polar angle, there is no 
conservation of the $z$ component of the angular momentum. In figures \ref{fig3}(a)--(b) 
we compare the orbits for the triaxial potential (solid curves) and for the axisymmetric potential 
(dashed curves). The parameters used were the same as in figure \ref{fig1} and for the 
axisymmetric case $\bar{a}_1=\bar{a}_2=0$. In the equations of motion time was 
rescaled as $\bar{t}=(Gm/a_3^3)^{1/2}t$ and the conserved mechanical energy per unit mass 
$E$ was rescaled as $\bar{E}=(Gm/a_3)^{-1}E$. Initial conditions in cylindrical coordinates 
were set as $\bar{R}=3.0$, $\dot{\bar{R}}=0$, $\varphi=0$ and $\dot{\varphi}$ was determined 
for a total energy $\bar{E}=-0.15$ in figure \ref{fig3}(a) and $\bar{E}=-0.10$ in figure 
\ref{fig3}(b). In the axisymmetric case the orbits shown are precessing ellipses. When 
axisymmetry is destroyed, so is the regular oscillation of the radial coordinate. At a higher 
energy, both orbits are qualitatively almost identical. 
\begin{figure}
\centering
\includegraphics[scale=0.68]{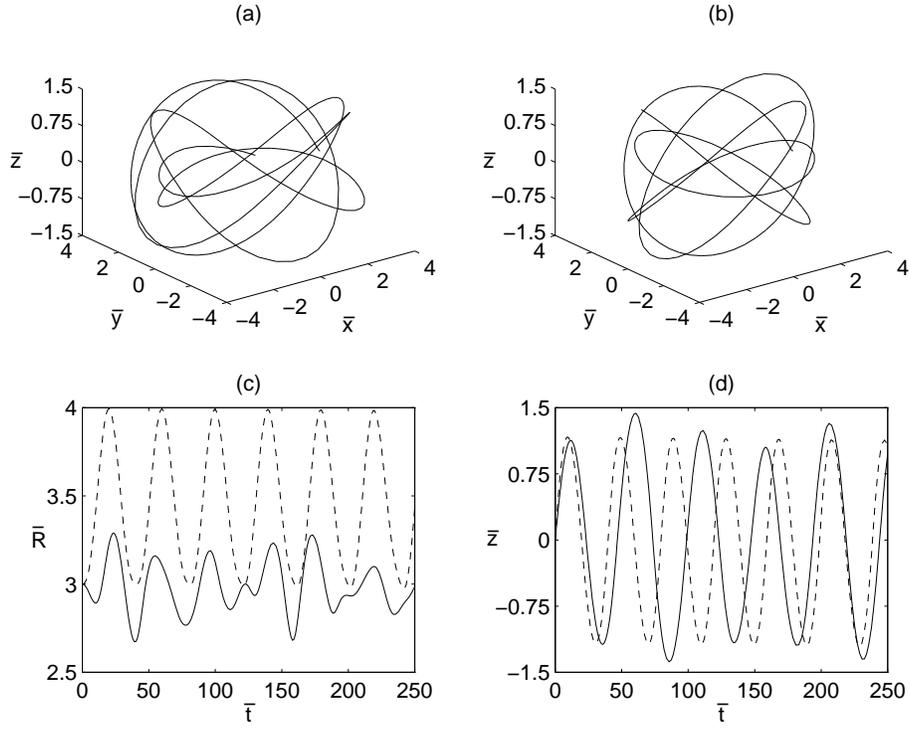}
\caption{Three-dimensional orbit for (a) axisymmetric potential and (b) triaxial potential. 
(c) Radial coordinate $\bar{R}$ as function of time $\bar{t}$. (d) Coordinate $\bar{z}$ as function 
of time $\bar{t}$. Solid curves represent motion for triaxial potential and dashed curves the 
axisymmetric case. The parameters are the same as in figure \ref{fig3}. Initial conditions: 
$\bar{R}=3.0$, $\dot{\bar{R}}=0$, $\varphi=0$, $\dot{\bar{z}}=0.5\bar{R}\dot{\varphi}$ and 
energy $\bar{E}=-0.15$.} \label{fig4}
\end{figure} 

Figures \ref{fig4}(a)--(d) show an example of three-dimensional orbit for 
(a) axisymmetric potential and (b) triaxial potential. In figures \ref{fig4}(c)--(d) the 
radial coordinate $\bar{R}$ and the coordinate $\bar{z}$, respectively, are plotted 
as function of time $\bar{t}$ for the triaxial case (solid curves) and  axisymmetric case 
(dashed curves). Here the parameters used were the same as in the previous example; the 
initial conditions were  $\bar{R}=3.0$, $\dot{\bar{R}}=0$, $\varphi=0$ and the imposed initial constrait 
 $\dot{\bar{z}}=0.5\bar{R}\dot{\varphi}$ was calculated for an energy of $\bar{E}=-0.15$. 
As before, motion with triaxial potential is more irregular compared with the axisymmetric one. 
This reflects  the loss of one of the motion integrals, the angular momentum.  
\section{Discussion} \label{sec_discuss}

By applying a Miyamoto and Nagai transformation on the three Cartesian coordinates of the monopole 
potential, we obtained a triaxial potential-density pair whose mass density distribution is 
everywhere non-negative and free from singularities; also, it is box-shaped with respect to the 
$z$ axis. A triaxial version of one of Satoh's axisymmetric models also yields a 
mass-density distribution with similar characteristics. We believe that these simple analytical
models may be useful for disk galaxies having box-shaped bulges. Some numerically calculated 
orbits for the Miyamoto and Nagai-like triaxial potential were also presented. 

\bigskip
D.\ Vogt thanks FAPESP for financial support. P.\ S.\ Letelier thanks CNPq and FAPESP for financial support. This research has 
made use of NASA's Astrophysics Data System.


\begin{thebibliography}{99}
\bibitem{Jaffe83} W. Jaffe, \textit{Mon. Not. R. Astron. Soc.} \textbf{202}, 995 (1983).
\bibitem{Hern90} L. Hernquist, \textit{Astrophys. J.} \textbf{356}, 359 (1990).
\bibitem{Miy75} M. Miyamoto and R. Nagai, \textit{Publ. Astron. Soc. Japan} \textbf{27}, 533 (1975).
\bibitem{Satoh80} C. Satoh, \textit{Publ. Astron. Soc. Japan} \textbf{32}, 41 (1980).
\bibitem{Zeeuw88} T. de Zeeuw and D.  Pfenniger, \textit{Mon. Not. R. Astron. Soc.} \textbf{235}, 949 (1988).
\bibitem{Long92} K. Long and C.  Murali, \textit{Astrophys. J.}  \textbf{397}, 44 (1992). 
\bibitem{Binney} S. Binney and S. Tremaine, \textit{Galactic Dynamics}, (Princeton University Press, Princeton, 1987).
\bibitem{Morgan69} T. Morgan and L.  Morgan, \textit{Phys. Rev.} \textbf{183}, 1097 (1969).
\bibitem{Morgan70} L. Morgan and T.  Morgan, \textit{Phys. Rev. D} \textbf{2}, 2756 (1970).
\bibitem{Bicak93} J. Bi\v{c}\'{a}k, D. Lynden-Bell and J.  Katz, \textit{Phys. Rev. D} \textbf{47}, 4334 (1993).
\bibitem{Lemos94} J. P. S. Lemos and P. S.  Letelier, \textit{Phys. Rev. D} \textbf{49}, 5135 (1994).
\bibitem{Gonzalez1} G. A. Gonz\'{a}lez and P. S.  Letelier, \textit{Phys. Rev. D} \textbf{62}, 064025 (2000).
\bibitem{Gonzalez2} G. A. Gonz\'{a}lez and P. S. Letelier, \textit{Phys. Rev. D} \textbf{69}, 044013 (2004).
\bibitem{Vogt1} D. Vogt and P. S.  Letelier, \textit{Phys. Rev. D} \textbf{68}, 084010 (2003).
\bibitem{Vogt2} D. Vogt and P. S.  Letelier, \textit{Phys. Rev. D} \textbf{71}, 084030 (2005). 
\bibitem{Vogt3} D. Vogt and P. S.  Letelier, \textit{Mon. Not. R. Astron. Soc.} \textbf{363}, 268 (2005).
\bibitem{Luetticke1} R. L\"utticke, R. J.  Dettmar and M. Pohlen,  \textit{Astron. \& Astrophys. S.} \textbf{145}, 405 (2000).
\bibitem{Luetticke2} R. L\"utticke, M. Pohlen and R. J. Dettmar, \textit{Astron. \& Astrophys.} \textbf{417}, 527 (2004).
\end{thebibliography}
\end{document}